\documentclass[letterpaper, 10pt, conference]{ieeeconf}      

\IEEEoverridecommandlockouts                              

\usepackage{graphicx}
\usepackage{epstopdf}
\usepackage{amsmath} 
\usepackage{amssymb}  
\usepackage[capitalise]{cleveref}
\usepackage{pifont}
\newcommand{\cmark}{\ding{51}}%
\usepackage{booktabs} 
\usepackage{multirow}
\newcommand{\ra}[1]{\renewcommand{\arraystretch}{#1}}
\usepackage{textcomp}
\usepackage{bm}
\usepackage{cite}
\usepackage[dvipsnames,svgnames,x11names]{xcolor}
\usepackage[markup=underlined]{changes}

\definechangesauthor[color=NavyBlue]{PM}
\definechangesauthor[color=BrickRed]{VS}

\newcommand{\var}{\text{var}}

\usepackage{etoolbox}
\makeatletter
\patchcmd{\@makecaption}
  {\scshape}
  {}
  {}
  {}
\makeatother

\title{\LARGE \bf
Promise and Challenges of a Data-Driven Approach for Battery Lifetime Prognostics
}

\author{Valentin Sulzer$^{1,\dagger}$, Peyman Mohtat$^{1}$, Suhak Lee$^{1}$, Jason B. Siegel$^{1}$, and Anna G. Stefanopoulou$^{1}$
\thanks{Distribution A. Approved for public release; distribution unlimited. (OPSEC 4603)}
\thanks{$^{1}$Department of Mechanical Engineering, University of Michigan, Ann Arbor}%
\thanks{$\dagger$
        {\tt\small vsulzer@umich.edu}}%
}

\begin{document}

\maketitle
\thispagestyle{empty}
\pagestyle{empty}

\begin{abstract}
Recent data-driven approaches have shown great potential in early prediction of battery cycle life by utilizing features from the discharge voltage curve. However, these studies caution that data-driven approaches must be combined with specific design of experiments in order to limit the range of aging conditions, since the expected life of Li-ion batteries is a complex function of various aging factors. In this work, we investigate the performance of the data-driven approach for battery lifetime prognostics with Li-ion batteries cycled under a variety of aging conditions, in order to determine when the data-driven approach can successfully be applied. Results show a correlation between the variance of the discharge capacity difference and the end-of-life for cells aged under a wide range of charge/discharge C-rates and operating temperatures. This holds despite the different conditions being used not only to cycle the batteries but also to obtain the features: the features are calculated directly from cycling data without separate slow characterization cycles at a controlled temperature. However, the correlation weakens considerably when the voltage data window for feature extraction is reduced, or when features from the charge voltage curve instead of discharge are used. As deep constant-current discharges rarely happen in practice, this imposes new challenges for applying this method in a real-world system. 

\end{abstract}

\section{INTRODUCTION}

The proliferation of Li-ion batteries is underway with a shift to electric vehicles and an increasing demand for commercial and residential energy storage systems. A key challenge is the prediction of cycle life under various usage patterns and operating temperatures. In particular, predictions that use only early-cycle data, without long historical information, can open a new chapter in battery design, production, and usage optimization \cite{attia2020closed}. However, early prediction is typically challenging because of the non-linearity of battery degradation. For instance, a weak correlation ($\rho$=0.1) was found in \cite{harris2017failure} between remaining capacity at cycle 80 and capacity values at cycle 500.

Approaches for prediction of the future degradation and cycle life of Li-ion batteries can be classified into three general categories: physics-based modeling of the main degradation mechanisms \cite{wang2011cycle,reniers2019review}, phenomenological modeling of capacity fade or internal resistance increase \cite{goebel2008prognostics, rezvanizaniani2014review}, and recently data-driven machine learning methods \cite{richardson2017gaussian, li2019data, severson2019data}.  

The physics-based degradation modeling approach uses partial differential equations (PDEs) to quantify the physical and chemical effects occurring inside a battery, such as ion diffusion and electrochemical reactions. 
This can then be used to explain input and output relationships, such as between input current profile and resulting capacity fade or resistance increase. 
This approach gives very detailed information about the processes occurring inside the battery, which can be used not only to predict degradation but also mitigate it through battery design and management.
However, the complexity of the models (several PDEs) and relative paucity of available data (current, voltage and temperature data) makes it hard to verify the accuracy, both of the model itself and of its parameterization.

The phenomenological modeling approach simplifies the prediction of degradation of Li-ion batteries by focusing on changes in the specific measures of degradation, such as internal resistance or cell capacity, with respect to cycle number or Amp-hour (Ah) throughput. 
For example, Goebel et al. \cite{goebel2008prognostics} use an exponential growth model as a degradation model to capture the increasing trend of internal resistance of the battery with respect to time in weeks. In this approach, parameters of the empirical degradation model (e.g. exponential growth model) are estimated from historical/previous data, and future trend is extrapolated at the time of making a prediction. 
This approach is appealing due to its simplicity, but it can fail to account for the complexity of Li-ion battery degradation, which usually depends on more than just time and cycle number, and entirely ignores the rich data available from the voltage curve.
Furthermore, since identifying the degradation model parameter depends on the available previous data, it often requires long historical data (at least 25\% along the trajectory to end-of-life \cite{zhang2018long}) to forecast future trends accurately. 

Recently, Severson et al. \cite{severson2019data} demonstrated a third approach that can be used to make predictions using only early-cycle data without the need for complex electrochemical models by showing that carefully extracted features from the discharge voltage curves in early (first 100) cycles can be used to predict a battery's cycle life, ranging from 150 to 2300 cycles, with 9.1\% test error. In their study, a dataset of 124 LFP/graphite A123 cells was generated from different fast charging rates and their charging profiles. The cells were cycled between full charge and full discharge with identical nominal temperatures and discharge C-rates, but varying charge C-rates. 


In this work, we construct a similar dataset of 12 NMC/graphite cells cycled to failure (as described in Section II) and probe the feasibility of this particular data-driven approach. Firstly, in Section III, we investigate whether the same features from the discharge voltage curves in early cycles show good correlation with cycle life, even for a different chemistry and with a broader range of operating temperatures and discharge C-rates, and show that end-of-life can be predicted with 16\% error.

Secondly, in Section IV, we explore whether the approach can be pushed further, beyond using full, constant-current discharges. To do this, we consider two scenarios in which full, constant-current discharge data could be unavailable. In the first scenario, constant-current discharge data is available but only for a partial data window. We examine how large the partial data window must be in order to provide adequate results. In the second scenario, no constant-current discharge data is available, only constant-current charge data. This scenario more closely mirrors a real-life situation in which discharges occur in response to user demand (for example, driving an electric vehicle), but charges can be more closely controlled. In this case, we investigate whether features in the charge voltage data can be used to predict cycle life.






\section{DATA GENERATION}

\begin{figure*}[ht]
    \centering
    \includegraphics[width=\linewidth]{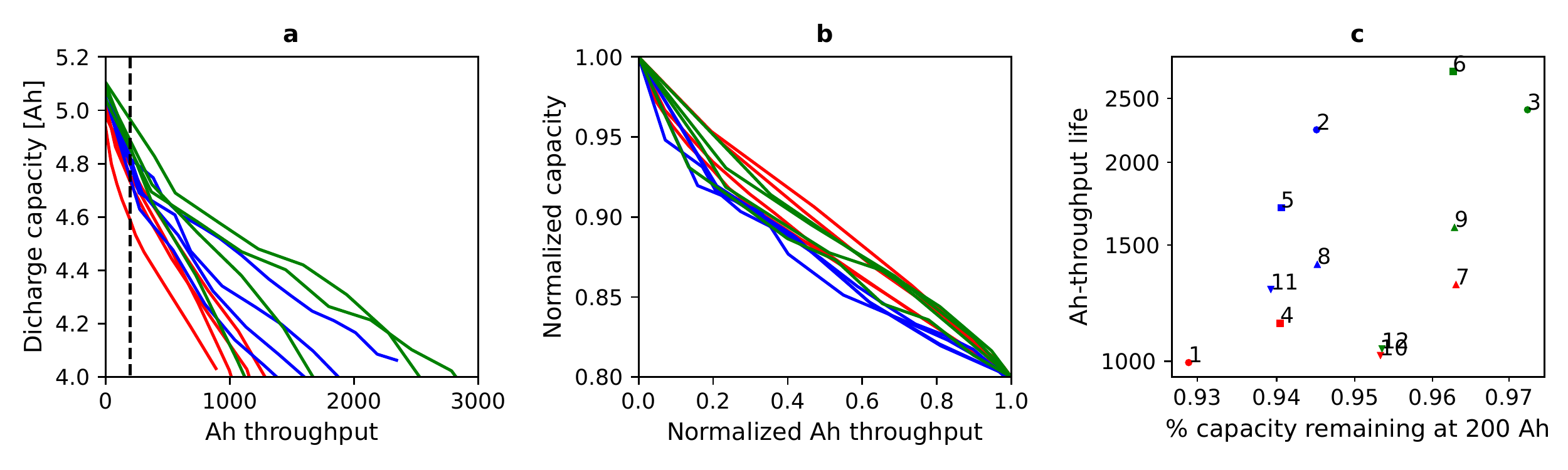}
    \caption{Discharge capacity curves and correlation with Ah-throughput life. (a) Discharge capacities as a function of Amp-hour (Ah) throughput. The dashed black line is at 200 Ah throughput. (b) Same data, but normalized so that 80\% relative capacity is reached at a normalized Ah throughput of 1, showing the relative shapes of the discharge capacity curves. (c) Correlation of percent remaining capacity after 200 Ah of throughput to Ah-throughput life. The correlation is 0.50. Colors denote temperature: hot (red), room (green), cold (blue).}
    \label{fig:discharge_capacities}
\end{figure*}

\begin{table}[!b]\centering
\ra{0.9}
\scriptsize
\begin{tabular}{p{4cm}p{3cm}}\toprule
{Pouch cell} & \\ \midrule
{Nominal capacity} & \multicolumn{1}{r}{5.0 Ah}\\
{Operating voltage} & \multicolumn{1}{r}{3.0-4.2 V}\\ 
{Thickness} & \multicolumn{1}{r}{4.0 mm}\\ 
{Length} & \multicolumn{1}{r}{132 mm}\\ 
{Width} & \multicolumn{1}{r}{90 mm}\\ \\
Positive electrode \\ \midrule
Material & \multicolumn{1}{r}{NMC111:CB:PVDF (94:3:3)}\\
\\
Negative electrode \\ \midrule
Material & \multicolumn{1}{r}{Graphite:PVDF (95:5)}\\
\\ 
Separator \\ \midrule
Material & \multicolumn{1}{r}{Polyethylene (PE)}\\ \\
Electrolyte \\ \midrule
Material & \multicolumn{1}{r}{1 M $LiPF_6$}\\ 
Organic solvent in electrolyte & \multicolumn{1}{r}{$2\%$ EC:EMC (3:7)}\\ \bottomrule
\end{tabular}
\caption{The pouch cell specifications.}
\label{table:pouch}
\end{table}


\begin{table*}[tbh]\centering
\ra{0.9}
\scriptsize
\begin{tabular}{ccccccccccccc}\toprule
\multicolumn{1}{l}{Cyclic aging conditions} &  &  &  &  &  &  &  &  &  &  &  &   \\ \midrule
\multirow{2}{*}{Condition group} & \multirow{2}{*}{Test number} & \multicolumn{3}{c}{Temperature$^a$} & \multicolumn{2}{c}{DOD} & \multicolumn{3}{c}{Charge$^b$} & \multicolumn{3}{c}{Discharge}  \\  \cmidrule(lr){3-5} \cmidrule(lr){6-7} \cmidrule(lr){8-10} \cmidrule(lr){11-13}
 &  & H & C & R & 0-100\% & 0-50\% & C/5 & 1.5C & 2C & C/5 & 1.5C & 2C   \\ \midrule 
A & 01, 02, 03 & \cmark & \cmark & \cmark & \cmark &  & \cmark &  &  & \cmark &  &   \\ 
B & 04, 05, 06 & \cmark & \cmark & \cmark & \cmark &  & \cmark &  &  &  & \cmark &   \\
C & 07, 08, 09 & \cmark & \cmark & \cmark & \cmark &  &  & \cmark &  &  & \cmark &   \\
D & 10, 11, 12 & \cmark & \cmark & \cmark & \cmark &  &  &  &  \cmark &  &  & \cmark  \\
\bottomrule
\end{tabular}
\caption{The aging test conditions matrix. $^a$ The H, C, and R corresponds to hot ($45^{\circ}C$), cold ($-5^{\circ}C$), and room ($25^{\circ}C$) temperature. $^b$ Constant current until 4.2 V and then constant voltage until ($I<C/50$).}
\label{table:conditions}
\end{table*}
\subsection{Experimental method}
\label{Experimental}
To study the degradation under a variety of conditions, 12 identical 5 Ah NMC/graphite pouch cells were selected from a batch manufactured at the University of Michigan Battery Lab (UMBL), with specifications shown in \cref{table:pouch}. Initial formation cycles were performed after manufacture to ensure the safety and performance stability of the cells. Then the cells were assembled inside fixtures. The dynamic tests were performed using a battery cycler (Biologic, France). The fixtures were installed inside a climate chamber (Cincinnati Ind., USA) in order to control the temperature during cycling and the characterization tests. The temperature was measured using a K-type thermocouple (Omega, USA) place on the surface of the battery.

The cycle aging experiments were designed to cover an array of test conditions such as different charge/discharge C-rates, and different nominal temperatures. The test conditions range from low C-rate room temperature baseline aging to high C-rate hot temperature accelerated aging, and are sumarized in \cref{table:conditions}. Each of the test conditions was performed at three different temperatures: hot ($45^{\circ}C$), cold ($-5^{\circ}C$), and room temperature ($25^{\circ}C$). Before the start of the cycling the cells were held at the target temperature for 3 hours to ensure thermal equilibrium. The cycling consists of a constant current (CC) charge until reaching 4.2 V, followed by a constant voltage (CV) phase at 4.2 V until the current falls below C/50, and finally CC discharge until reaching 3.0 V.

Intermediate diagnostic (C/20) tests were performed at a periodic number of cycles corresponding to an expected 5\% capacity loss.
For these diagnostic tests, the cells were brought back to the room temperature ($25^{\circ}C$) and held at rest for 3 hours to ensure thermal equilibrium. 
The diagnostic test consists of an initial C/5 discharge until reaching 3.0 V, followed by a constant voltage (CV) phase at 3.0 V until the charge current falls below C/50 and 1 hour rest to ensure the cell is fully discharged. This is followed by a C/20 charge until reaching 4.2 V, then a constant voltage (CV) phase at 4.2 V until until the charge current falls below C/50 and 1 hour rest. Finally, the cell is discharged at C/20 until reaching 3.0 V.



\subsection{Discharge capacity data}

The discharge capacity curves of the 12 cells cycled to full depth-of-discharge are shown in \Cref{fig:discharge_capacities}a.
We use the discharge capacity as measured by the diagnostic tests, since the cells do not reach the capacity limits during cycling due to high internal resistance, especially at cold temperatures and/or high C-rates.

Discharge capacity is plotted as a function of Ah-throughput, rather than cycle number, because cells charged and discharged between fixed voltage limits at different temperatures and C-rates observe different Ah-throughput per cycle.
For these 5 Ah cells, an upper bound for one equivalent cycle is 10 Ah of throughput (including charge and discharge).
We define ``Ah-throughput life'' as the Ah-throughput at which 80\% of initial capacity was reached; this is the equivalent of cycle life when measuring by cycle number.

The wide variety of operating conditions gives rise to a wide range of Ah-throughput lives for the cells, ranging from 1000 Ah to 3000 Ah.
Unlike the LFP/graphite cells used by Severson et al. \cite{severson2019data}, the NMC/graphite cells used here do exhibit significant capacity loss during early cycling.
\Cref{fig:discharge_capacities}c shows that the percent capacity remaining after the first 200 Ah of throughput does correlate with Ah-throughput life, but only weakly (Pearson correlation coefficient of 0.50).
A similar weak correlation between early capacity loss and lifetime capacity loss has previously been reported in the literature \cite{harris2017failure}.
The correlation of the early percentage capacity loss with Ah-throughput life provides a baseline benchmark for features in the discharge voltage curve.

On average, the cells cycled at hot temperature degraded the fastest, followed by the cells cycled at cold temperature, and the cells cycled at room temperature degraded the slowest.
Meanwhile, higher charge and discharge rates led to faster degradation on average.
While these observations hold in an averaged sense, there were also significant outliers. For example, cell 1, cycled at low C-rate, and cell 12, cycled at room temperature, were among the fastest degrading cells. 

The fact that averaged trends were as expected but with significant outliers suggests that differences in degradation rate were driven by a mix of operating variations and manufacturing variations.
It should be notes that the cells used in these experiments were designed as energy cells, rather than power cells, hence the poor performance at these relatively high C-rates.


\section{MACHINE LEARNING APPROACH}

\subsection{Features extraction from discharge voltage curves}

\begin{figure*}[!ht]
    \centering
    \includegraphics[width=\linewidth]{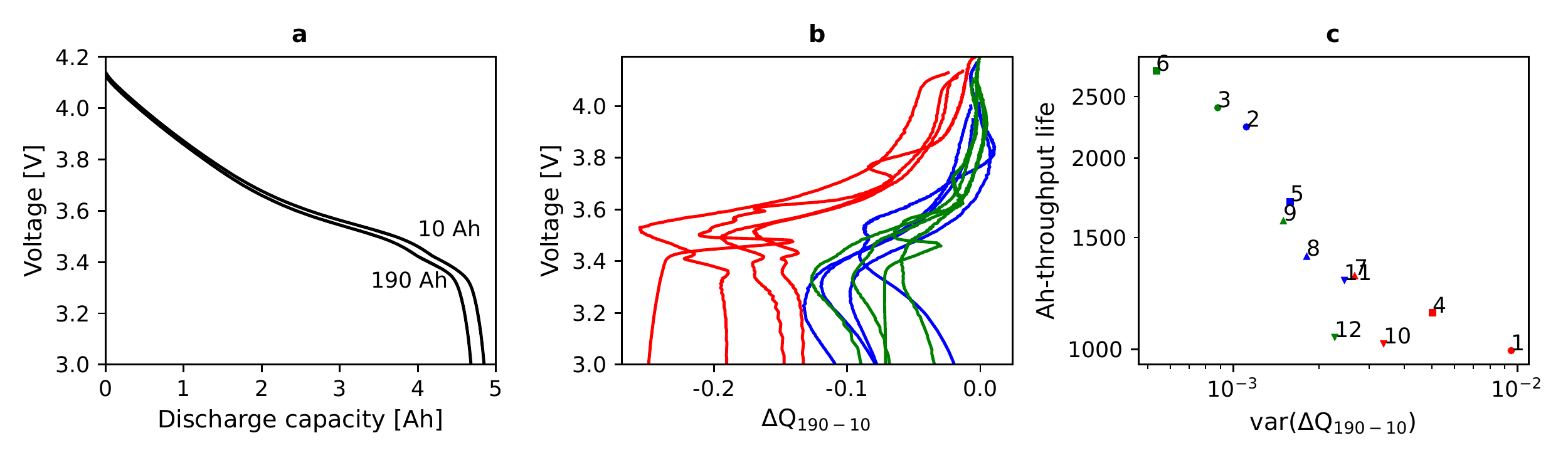}
    \caption{Features from the discharge voltage curves. (a) Typical discharge voltage curves after 10 Ah and 190 Ah of cycling. (b) Difference between discharge voltage curves from cycles taken after 10 Ah of cycling and 190 Ah of cycling. (c) Ah-throughput life against variance of $\Delta Q_{190-10}$, plotted on  a log-log scale. The correlation is $-0.90$.}
    \label{fig:DeltaQ_discharge}
\end{figure*}

Charge and discharge voltage curves contain much more information than simply providing the capacity of the cell. Typically, state-of-health-related information is extracted from either a charge/discharge voltage curve or its derivative with respect to discharge capacity (differential voltage analysis), on a cycle-by-cycle basis. For example, electrochemical models can be parametrized using voltage curves, or features can be directly extracted from them \cite{birkl2017degradation, lee2020li}. 

Severson et al. \cite{severson2019data} propose a new approach, comparing two discharge voltage curves from different cycles, shown here in \Cref{fig:DeltaQ_discharge}a.
By inverting the discharge voltage curve to find the discharge capacity as a function of voltage, and then taking the difference in discharge capacities between cycles, we obtain discharge capacity difference
\begin{equation}
    \Delta Q_{x-y}(V) = Q_x(V) - Q_y(V).
    \label{eq:1}
\end{equation}
Here, we define $Q_x$ to be the discharge capacity as a function of voltage for the cycle in which $x$ Ah throughput was reached. The typical convention is to use $x>y$, so that $\Delta Q_{x-y}$ is negative, but this is not strictly necessary.

\begin{figure}
    \centering
    \includegraphics[width=\linewidth]{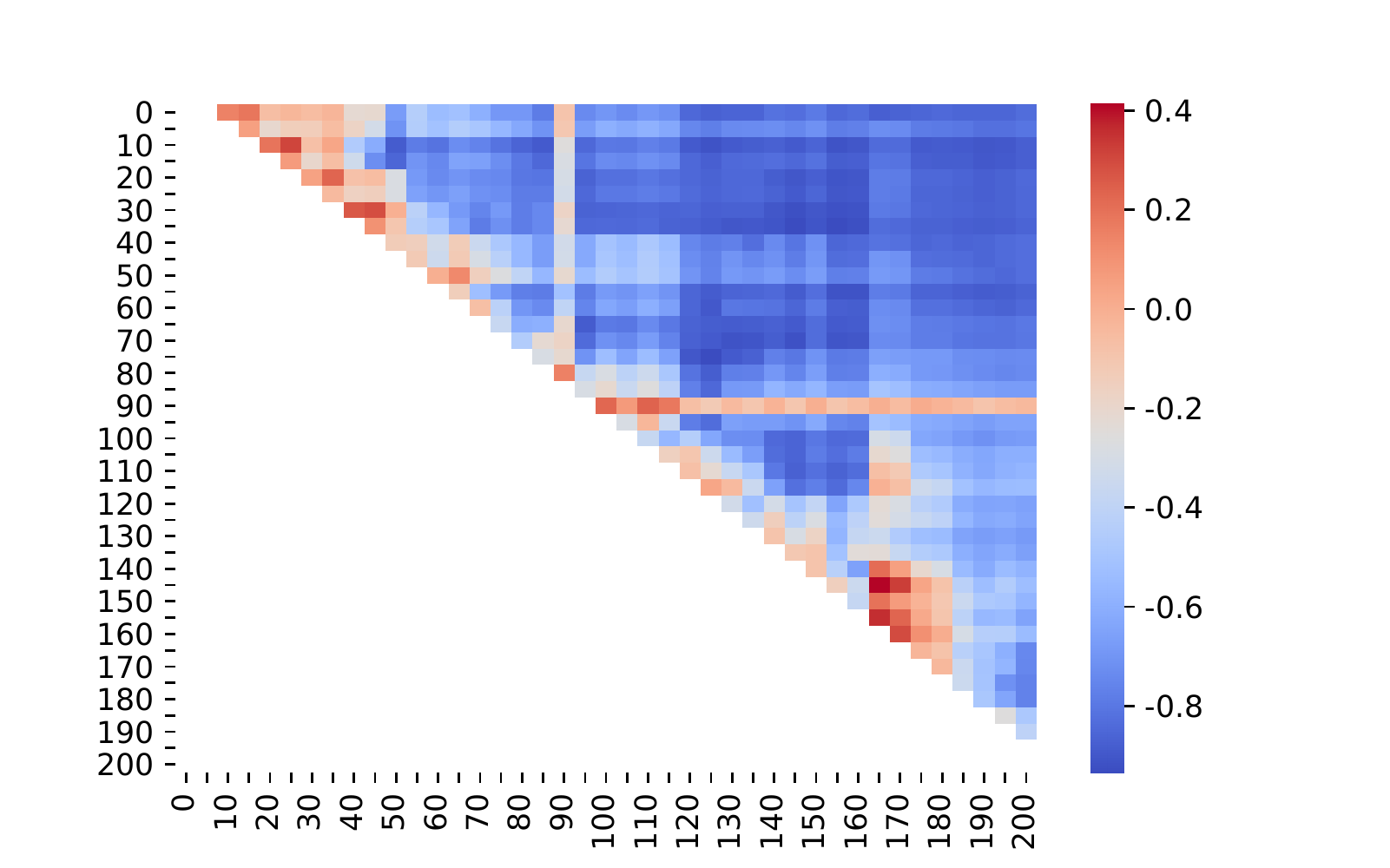}
    \caption{Heatmap of the log-correlation of the variance of $\Delta Q_{x-y}$ with Ah-throughput life when changing the Ah throughput values used to calculate $\Delta Q$. In general, a larger difference between $x$ and $y$ gives better correlation.}
    \label{fig:heatmap}
\end{figure}
All high and low combinations of Ah throughput values are systematically tested (\Cref{fig:heatmap}), restricting to below 200 Ah to verify the early-prediction capability. From this analysis, we find that the values of 10 Ah and 190 Ah give the best log-correlation between the variance of $\Delta Q$ and the Ah-throughput life. In general, the correlation is stronger when using cycles with a larger difference between high and low values for the Ah throughput, which gives further confidence that the variance of the $\Delta Q$ curve is capturing physically relevant effects. 

\Cref{fig:DeltaQ_discharge}b shows $\Delta Q_{190-10}$ for all 12 cells, and \Cref{fig:DeltaQ_discharge}c shows the variance of each of these curves plotted against Ah-throughput life on a log-log axis, with a strong negative correlation (correlation of -0.90).
Other statistics of the $\Delta Q_{190-10}$ curve, such as the minimum and mean values, also show very good correlation, while the skew and kurtosis show poor correlation (see \Cref{table:discharge_stats}).

\begin{table}[!t]\centering
\ra{1}
\begin{tabular}{cc}
\toprule
Statistic & Log-correlation \\
\midrule
Mimimum & -0.87 \\
Mean & -0.84 \\
Variance & -0.90 \\
Skew & -0.18 \\
Kurtosis & -0.24 \\
\bottomrule
\end{tabular}
\caption{Log-correlation of various statistics of the discharge difference curve, $\Delta Q_{190-10}$, with Ah-throughput life.}
\label{table:discharge_stats}
\end{table}

This excellent correlation holds despite the wide range of operating temperatures and charge/discharge C-rates that were used not only to cycle the cells but also to obtain the $\Delta Q_{190-10}$ curve (the $\Delta Q_{190-10}$ curve is calculated directly from cycling data, rather than from separate low C-rate characterization cycles at room temperature).

\subsection{Linear regression model for prognostics}

To demonstrate the effectiveness of using features from the discharge voltage curve for battery lifetime prognostics, we use some of the data to train a simple machine learning algorithm, then evaluate its predictive power on a test set.

We restrict ourselves to simple regularized linear regressions as the small size of the dataset could easily lead to over-fitting if using more advanced algorithms.
The focus of this work is to better understand the features themselves, and what are the scenarios in which they can be useful for prognostics, rather than optimization of the prognostics algorithm itself.
For larger datasets, more advanced algorithms such as the Relevance Vector Machine have been shown to be promising for this type of application \cite{fermin2020identification}.

Based on the correlations in \Cref{table:discharge_stats}, and for simplicity given the small amount of available data, we use the logarithms of variance, mean, and minimum of $\Delta Q_{190-10}$ as features, and the log of Ah-throughput life as the objective.
Note that final errors are reported for the Ah-throughput life (not its logarithm).

We use a regularized linear regression model, the Ridge regression model. With this algorithm, we find a vector of weights $\bm{w}^*$ that minimizes the cost function
\begin{equation}
    J(\bm{w}) = 
    \|\bm{y} - \bm{X}\bm{w}\|_2
    + \alpha\|\bm{w}\|_2
\end{equation}
where $\bm{y}$ is the vector of data and $\bm{X}$ is the matrix of features.
The hyperparamer $\alpha$ was tuned for optimal results, to a value of $8$.
We choose the ridge regression model because the model only has three features, all of which are known to correlate strongly with the objective, so L1-regularizations such as LASSO regression are not necessary (L1-regularization is useful when trying to obtain sparse models where some of the weights are set to zero).
We implement the ridge regression using the Python packages numpy \cite{numpy}, pandas \cite{reback2020pandas, mckinney-proc-scipy-2010}, and scikit-learn \cite{scikit-learn}.

\begin{figure}
    \centering
    \includegraphics[width=\linewidth]{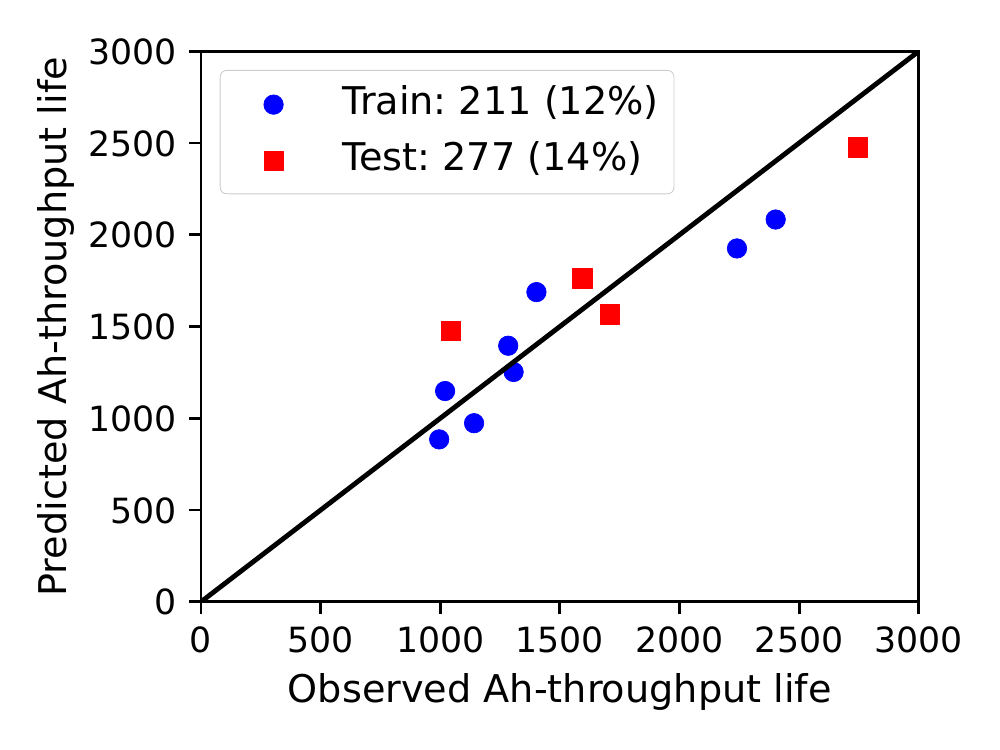}
    \caption{Ah-throughput life predictions using a ridge regression algorithm with the logarithms of minimum, mean, and variance of $\Delta Q_{190-10}$ as features. The legend shows root-mean-square error (in Ah) and mean percentage error. For comparison, a aseline prediction (from a regression to the mean) for this case gives 513 Ah training RMSE (27\%) and 720 Ah testing RMSE (39\%).}
    \label{fig:prediction}
\end{figure}

We randomly split the data into training and testing sets with an 8/4 ratio.
Averaging over 100 such random splits, the average training RMSE is 269 Ah (13\% MPE) and the average testing RMSE is 335 Ah (16\% MPE).
This can be compared to the average error from a simple regression to the mean, which gives an average training RMSE of 539 Ah (29\%) and an average testing RMSE of 617 Ah (35\%).
Therefore, using features from the discharge voltage curve approximately halves the error of lifetime prediction.
In \Cref{fig:prediction}, we show cell-by-cell predictions for a single representative train/test split, which gives a training RMSE of 211 Ah (12\% MPE) and a testing RMSE of 277 Ah (14\% MPE). 

\section{DISCUSSION}

\subsection{Performance using discharge voltage features}

The strong correlation of the features obtained from the discharge voltage curve verifies the capability of the presented data-driven approaches even for different cell chemistry (LFP/graphite in \cite{severson2019data} and NMC/graphite in this study), a wide range of operating temperatures, and various C-rates for charge/discharge cycling.  

\begin{figure*}
    \centering
    \includegraphics[width=\linewidth]{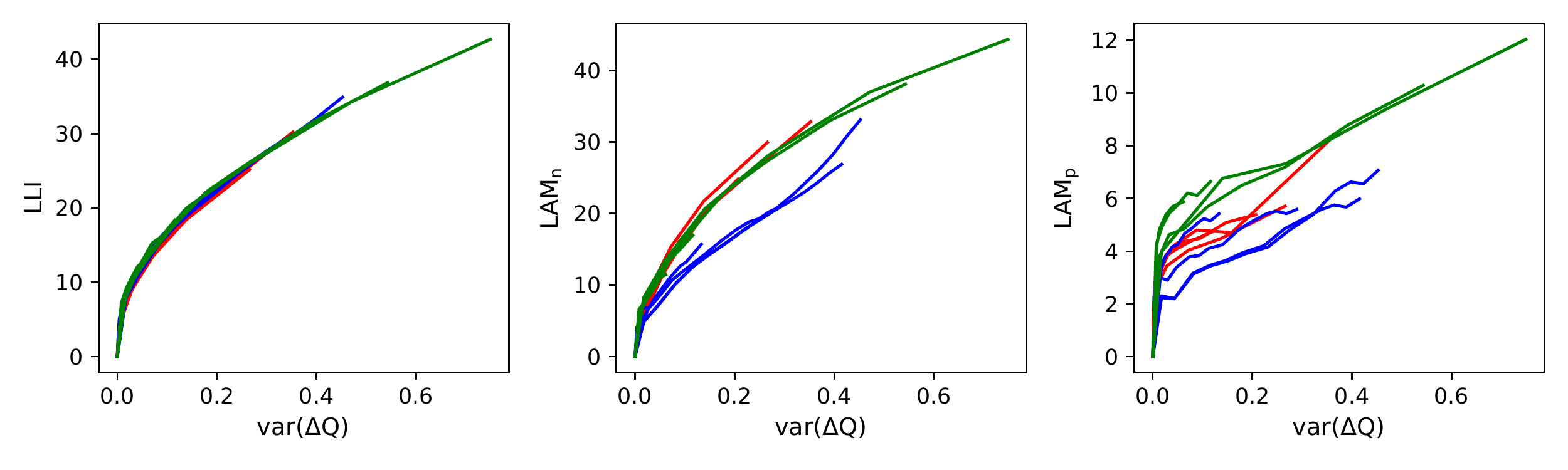}
    \caption{Map between $\var(\Delta Q)$ and the degradation mechanisms of loss of lithium inventory (LLI), loss of active material in the negative electrode (LAM$_n$), and loss of active material in the positive electrode (LAM$_p$).}
    \label{fig:LLI_LAM}
\end{figure*}
In the case of these NMC/graphite cells, the degradation mechanism leading to these features is also visible in the discharge capacity curve even after 200 Ah of cycling, but using the feature from the discharge voltage curve gives superior predictive power. Using differential voltage analysis on the characterization tests, we can better understand what are the degradation mechanisms occurring in these cells, and hence rationalize the predictive power of the discharge voltage features. 
For each characterization cycle (C/20 discharges at room temperature), we use the algorithm of Mohtat et al. \cite{mohtat2019towards} to determine which form of degradation has occurred: loss of lithium inventory (LLI), or loss of active material (LAM) in the negative or positive electrode. At the same time, we calculate $\var(\Delta Q)$ for that cycle by taking the difference in the discharge voltage curve with the first characterization cycle for that cell.
This allows us to build a map between $\var(\Delta Q)$ and LLI, LAM$_n$, and LAM$_p$, shown in \Cref{fig:LLI_LAM}.
These results show that $\var(\Delta Q)$ tracks LLI in exactly the same way at all C-rates and temperatures, which may be why $\var(\Delta Q)$ is such a good predictor of end-of-life.
Note that there may be other forms of degradation, such as SEI formation, lithium plating, or particle cracking, that also contribute to $\var(\Delta Q)$ but are not captured by LLI or LAM in the characterization cycles.


\begin{figure}
    \centering
    \includegraphics[width=\linewidth]
    {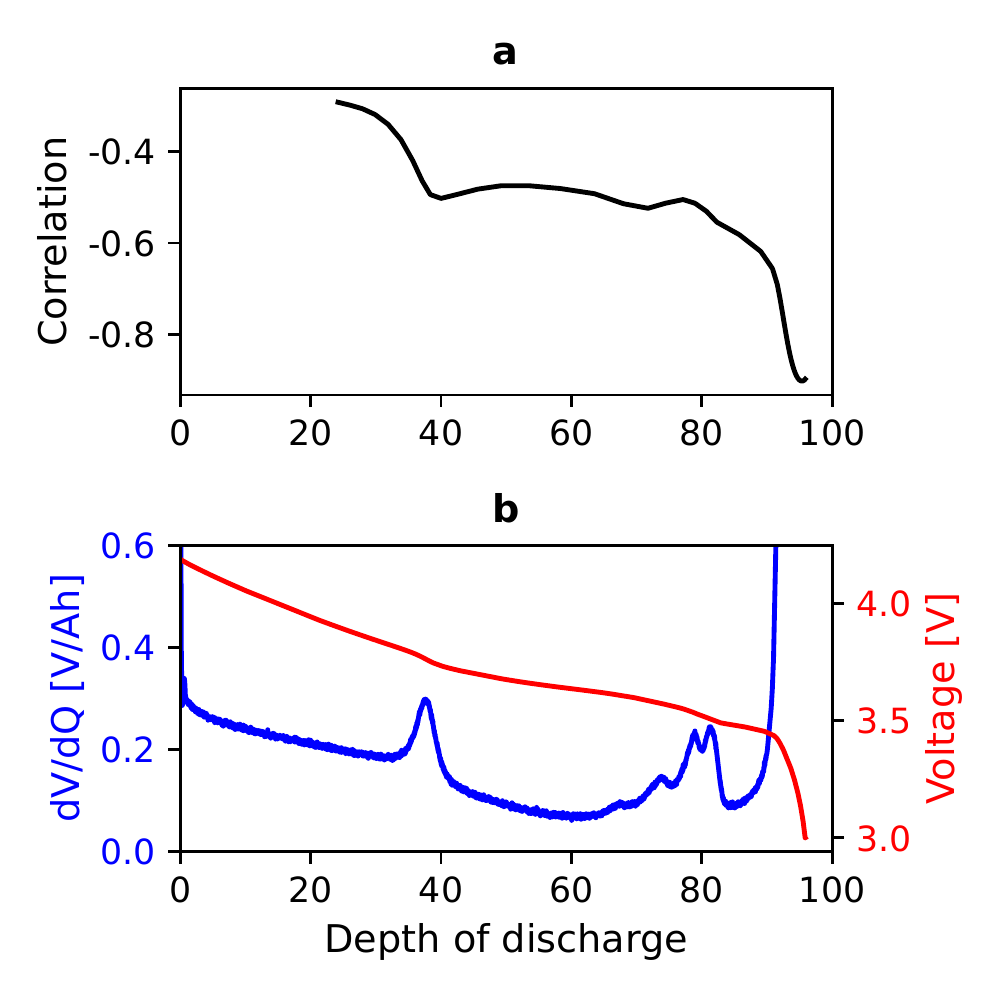}
    \caption{Correlation between the log of the variance of $\Delta Q_{190-10}$ and the Ah-throughput life, when varying the minimum voltage available to calculate $\Delta Q_{190-10}$. The bottom plot shows voltage, and differential voltage (dV/dQ), for reference.}
    \label{fig:effect_of_min_voltage}
\end{figure}

\subsection{Effect of partial data window}

In a real-life system, full discharges rarely occur, and so typically only partial discharge data is available.
We investigate whether this method is still useful in this case by reducing the size of the window used for the calculation of $\Delta Q_{190-10}$, and plotting the resulting correlation of its variance with Ah-throughput life (\Cref{fig:effect_of_min_voltage}).
There is a very rapid drop-off in the log-correlation between variance and Ah-throughput life as soon as the depth of discharge is reduced from 100\% (i.e. whenever the final voltage drop-off in \Cref{fig:discharge_capacities}a is not captured in $\Delta Q$).
Other features (minimum, mean) in the $\Delta Q_{190-10}$ curve show similar correlation with Ah-throughput life when reducing the size of the data window.
\Cref{fig:effect_of_min_voltage}b shows the voltage and differential voltage (dV/dQ) for a representative cell as a function of depth of discharge. This shows that it is important to capture the final voltage drop-off, between 90 and 100\% depth of discharge, to get the best correlation.
Furthermore, the correlation decreases further when the cell is not discharged below the peak in dV/dQ at 40\% depth of discharge.

These results suggest that, for NMC/graphite cells, this data-driven approach is only useful if full discharge data is available.
This effect is likely to also be significant for LFP/graphite cells used in \cite{severson2019data}, since the portion of the voltage curve that produces the $\Delta Q$ feature is only reached at around 90\% depth of discharge.


\subsection{Performance using charge voltage features}

\begin{figure*}
    \centering
    \includegraphics[width=\linewidth]{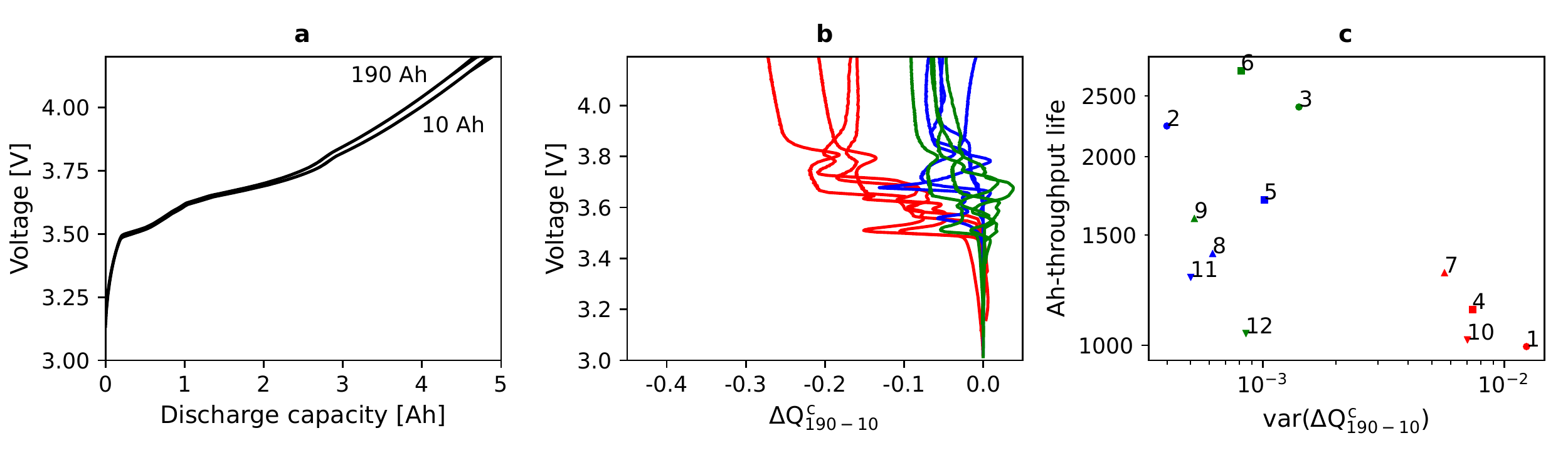}
    \caption{Features from the charge voltage curves. (a) Typical charge voltage curves after 10 Ah and 190 Ah of cycling. (b) Difference between charge voltage curves from cycles taken after 10 Ah of cycling and 190 Ah of cycling. (c) Ah-throughput life against variance of $\Delta Q^c_{190-10}$, plotted on  a log-log scale. The correlation is $-0.53$.}
    \label{fig:DeltaQ_charge}
\end{figure*}
In a real-life system, constant-current discharge voltage data is unlikely to be available since the cell's discharge current must continuously change in order to meet the user's power requirements.
Constant-current charge is more likely to be available, since CCCV (constant current, constant voltage) charging is the industry standard.

Considering this scenario, we investigate whether features from the constant-current charge voltage could instead be used for lifetime prediction.
The approach is identical to that used for features for the discharge difference curves: we take the difference of two constant-current charge curves, one after 10 Ah of throughput and one after 190 Ah of throughput, and then take various statistics of this curve.
\Cref{fig:DeltaQ_charge} shows this process, with \Cref{fig:DeltaQ_charge}c showing the correlation of the logarithms of $\Delta Q^c_{190-10}$ and Ah-througput life, with only weak negative correlation.
The log-correlations of other statistics for the charge voltage difference curve are similarly weak (\Cref{table:charge_stats}).
Hence the early-cycle prognostics algorithm does not perform as well when using features from the charge curves instead of the discharge curve.

One possible explanation for this reduced performance could be that some of the $\Delta Q^c_{190-10}$ curves, shown in \Cref{fig:DeltaQ_charge}, are missing the lower voltage range: as the cells degrade, the jump in voltage at the beginning of charge becomes much more significant, whereas this is not the case for discharge.
Hence the data window available for calculating the feature is reduced, and so the accuracy is reduced, as discussed above.


\begin{table}[!t]\centering
\ra{1}
\begin{tabular}{cc}
\toprule
Statistic & Log-correlation \\
\midrule
Mimimum & -0.43 \\
Mean & -0.44 \\
Variance & -0.57 \\
Skew & 0.35 \\
Kurtosis & 0.22 \\
\bottomrule
\end{tabular}
\caption{Log-correlation of various statistics of the charge difference curve, $\Delta Q^c_{190-10}$, with Ah-throughput life.}
\label{table:charge_stats}
\end{table}

\section{CONCLUSIONS}

This work verifies the capability of the data-driven prognostics in early-prediction of Li-ion battery lifetime using features from the discharge capacity difference curve in the case where the discharge C-rates and operating temperatures vary.
This suggests that the data-driven approach is very promising in the case where constant-current discharge data can be deliberately generated, even if the operating temperatures and discharge currents vary.


However, we also found that the suggested feature (the variance of discharge capacity difference) loses the strong correlation when using either partial discharge voltage data, or constant-current charge data (for example, as part of CCCV charging). In those cases, different features should be found that have better predictive power. In the case where full constant-current discharges are not available, one promising data-driven alternative is to use the constant-voltage phase of CCCV charging, since this data will almost always be available even in the cases of partial cycling and varying-current discharge. In particular, the time constant of the exponentially decaying current during the CV phase has been shown to be a good predictor of state-of-health \cite{eddahech2014determination, yang2018online, wang2019state}, and therefore may also be a good predictor of end-of-life.

This study opens the question of whether the range of degradation rates in the data is caused by the different operating conditions or manufacturing variability between the cells.
Investigating how this data-driven approach performs in either extreme (low variability, a wide range of conditions or high variability, narrow range of conditions) requires specifically generated data and remains an interesting open question.
The good performance of the algorithm for these NMC/graphite cells with reasonably high manufacturing variability (as the cells are used outside of their recommended conditions) suggests that the algorithm may be robust in both extremes.






\section*{ACKNOWLEDGMENTS}
\footnotesize{The experimental work in this material was supported by the Automotive Research Center (ARC) in accordance with Cooperative Agreement W56HZV-14-2-0001 U.S. Army CCDC GVSC. The authors would also like to thank the University of Michigan Battery Lab for providing the cells used to generate experimental data. Distribution A. Approved for public release; distribution unlimited (OPSEC 4603).}


\bibliographystyle{IEEEtran}
\bibliography{IEEEabrv,main}

\end{document}